\documentclass[10pt,onecolumn]{IEEEtran}
\usepackage[numbers,sort&compress]{natbib}
\usepackage{amsmath,amssymb,epsfig}
\usepackage{fancybox}
\usepackage{tikz}
\newtheorem{definition}{\bf Definition}
\newtheorem{theorem}{\bf Theorem}
\newtheorem{corollary}{\bf Corollary}

\begin{document}

\bibliographystyle{ieeetr}

\setlength{\parindent}{1pc}

\title{Statistical Properties of Loss Rate Estimators in  Tree Topology}
\author{Weiping~Zhu  \thanks{Weiping Zhu is with University of New South Wales, Australia, email w.zhu@adfa.edu.au}}
\date{}
\maketitle

\begin{abstract} Three types of explicit estimators are proposed here to estimate the loss rates of the links in a network of the tree topology. All of them are derived by the maximum likelihood principle and proved to be either asymptotic unbiased or unbiased. In addition,  a set of formulae are derived to compute the efficiencies and variances of the estimators that also cover some of  the estimators proposed previously. The formulae unveil that the variance of the estimates obtained by a maximum likelihood estimator for the pass rate of the root link of a multicast tree is equal to the variance of the pass rate of the multicast tree divided by the pass rate of the tree connected to the root link. Using the formulae, we are able to evaluate the estimators proposed so far and select an estimator for a data set.

\end{abstract}


\section{Introduction}
\label{section1}
Network characteristics, such as link-level loss rate, delay distribution, available
bandwidth, etc. are valuable information to network
operations, development and researches. Therefore,
 a considerable attention has been given to network
measurement, in particular to  large networks that cross a number of autonomous systems, where security concerns, commercial interests, and administrative
boundary make direct measurement impossible. To overcome the security and administrative obstacles, network tomography was
proposed in \cite{YV96}, where the author suggests the use of
end-to-end measurement and statistical inference to estimate the characteristics of interest.
Since then,
many works have been carried  out to estimate various characteristics that cover loss tomography \cite{CDHT99, CDMT99, CDMT99a, CN00, XGN06, BDPT02,ADV07, DHPT06, ZG05, GW03}, delay tomography
\cite{LY03,TCN03,PDHT02, SH03, LGN06},
 loss pattern tomography \cite{ADV07}, and so on. Despite the enthusiasm  in loss tomography, there has been little work to study the statistical properties of an estimator with a finite sample size although some asymptotic properties are presented in the literature \cite{CDHT99, DHPT06}.  The  finite sample properties, such as efficiency and variance,  differ from  the asymptotic ones that are critical to the performance evaluation of  an estimator since each of them unveil the quality and effectiveness of an estimator in a specific aspect.  Apart from that, the finite sample properties can be used to select a better estimator, if not the best,  from a group for a data set obtained from a specific circumstance. To fill the gap, we in this paper propose a number of maximum likelihood estimators (MLE) that can be solved explicitly for a network of the tree topology and provide  the statistical properties of them. The statistical properties are further extended to cover the MLEs proposed previously. One of the most important discoveries is a set of formulae to compute the efficiency and variance of the estimates obtained by an estimator.

The approach proposed in \cite{YV96} requires us to send probing packets, called probes, from some end-nodes called sources  to the
receivers located on the other side of the network, where the paths
connecting the sources to the receivers cover the links of interest. To make the probes received informative in statistical inference,
 multicast or unicast-based multicast proposed in \cite{HBB00,CN00}  is used to send probes from a source to a number of receivers, via a number of intermediate nodes that replicate the arrived probes and then forward to its descendants. This process continues until either the probes reach the destinations or lost, which makes the observations of any two receivers correlated in some degree and the degrees vary depending on the interconnection between the receivers.  Given the network topology used for sending probes and the observations obtained at receivers, we are able to create a likelihood function to connect the observation  to the process described above. 
 Since the number of correlations created by multicasting are proportional to the number of descendants attached to a node, the likelihood equation obtained for a node having many descendants is a high degree polynomial that requires an iterative procedure, such as  the expectation and maximization (EM) or the Newton-Raphson algorithm, to approximate the solution. Using iterative procedure to solve a polynomial has been widely criticised  for its computational complexity that increases with the number of descendants attached to the link or path to be estimated \cite{CN00}.  There has been a persistent effort in the research community to search for explicit estimators that are comparable in terms of accuracy to the estimators using iterative approach.  To achieve this, we must have the statistical properties of the estimates obtained by an estimator, such as unbiasedness, efficiency, and variance. Unfortunately, there has been little work in a general form for the properties and the asymptotic properties obtained in \cite{CDHT99, DHPT06} has little use in this circumstance.

  To overcome the problems stated above,   we have undertaken a thorough and systematic investigation of the estimators proposed for loss tomography that aims at identifying the statistical principle and strategies that have been used or can be used in the tree topology. A number of findings are obtained in the investigation that show all of the estimators proposed previously  rely on observed correlations to infer the loss/pass rates and most of them use all of the correlations available in estimation, such as the MLE  proposed in \cite{CDHT99}. However, the qualities of the correlations, measured by the fitness between a correlation and the corresponding observation, are very much ignored. Rather than using all of the correlations available in estimation, we propose here to use a small portion of high-quality ones and expect the estimates obtained by such an estimator are comparable to that considering all of the correlations.  The investigation further leads to a number of findings that contribute to loss tomography in four-fold.

\begin{itemize}
\item A large number of explicit estimators are proposed on the basis of composite likelihood \cite{Lindsay88} that are divided into three groups: the block wised estimators (BWE), the reduce scaled  estimators (RSE), and the individual based estimators (IBE).
\item The estimators in BWE and IBE are proved to be unbiased and that in RSE are proved to be  asymptotic unbiased as that proved in   \cite{DHPT06}. A set of formulae are derived for the efficiency and variances of the estimators in RSE and IBE, plus the MLE proposed in \cite{CDHT99}. The formulae show the variance of the estimates obtained by a MLE can be exactly expressed by the pass rate of the path of interest and the pass rate of the subtrees connected to the path.  The formulae also show the weakness of the result obtained in \cite{DHPT06}.
\item  The efficiency of the estimators in IBE are compared with each other on the basis of the Fisher information that  shows an estimator considering  a correlation involving a few observers can be more efficient than that considering more and the estimator proposed in \cite{DHPT06} is the least efficient. A similar conclusion is obtained for the estimators in BWE.
\item  Using the formulae, we able to identify an efficient estimator by examining the end-to-end observation that makes model selection not only possible but also feasible. A number of simulations are conducted to verify this feature that also show the connection between efficiency  and robustness of an estimator.
    \end{itemize}

The rest of the paper is organised as follows. In Section \ref{related work},  we briefly introduce the previous works related to explicit loss rate estimators and point out the weakness of them. In Section \ref{section2}, we introduce the loss model, the notations, and the statistics used in this paper.  Using the model and statistics, we  derive a MLE that considers all available correlations for a network of the tree topology in Section \ref{section3}. We then decompose the MLE into a number of components according to correlations and derive a number of likelihood equations for the components in Section \ref{section 4}. A statistic analysis of the proposed estimators is presented in Section \ref{section5} that details the statistical properties of the proposed estimators, one of them is the formulae to calculate the variances of various estimators. Simulation study is presented in Section \ref{section 6} that compares the performance of five estimators and shows the feasibility of selecting an estimator for a data set.
Section \ref{section7} is devoted to concluding remark.

\section{Related Works}\label{related work}

Multicast Inference of Network Characters (MINC) is the pioneer of using the ideas proposed in \cite{YV96} into practice, where a
Bernoulli model is used to model  the loss behaviors of a path. Using
this model, the authors of \cite{CDHT99} derive an estimator in the form of a polynomial that is one degree less than the number of descendants
connected to the end node of the path of interest \cite{CDHT99, CDMT99,CDMT99a}.  Apart from that, the authors obtain a number of results  from asymptotic theory, such as the large number behaviour of the estimator and the dependency of the estimator variance on topology. Unfortunately, the results only hold if the sample size $n$ grows indefinitely. In addition, if $n\rightarrow \infty$,  almost all of the estimators proposed previously must have the same results and no one can tell the difference between them.  In order to evaluate the performance of an estimator, experiments and simulation have been widely used but lead to little result since there are too many random factors affecting the results obtained from experiments and simulations.

To overcome the problem stated, simple and explicit estimators, such as that proposed in \cite{DHPT06}, are investigated that aims at reducing the complexity of an estimator and hopefully  finding theoretical support for further development since a simple estimator may be easy to analyse.
Using this strategy, the authors of \cite{DHPT06} propose an explicit estimator that only considers a correlation, i.e. the correlation involving all descendants, and claim the same asymptotic variance for the estimates obtained by the estimator as that obtained by the estimator proposed in \cite{CDHT99} to first
order. The claim is obtained by applying  the central limited theorem (CLT) on one of the results acquired by the asymptotic theory in \cite{CDHT99}, where  the covariance between two descendants attached to the path of interest is obtained by assuming the loss rate of a link is very small and then the delta method is used to compute the asymptotic variance on the covariance matrix obtained by the asymptotic theory. The repeated use of the CLT makes the claim questionable and expensive to use in practice since the result only holds if $n \rightarrow \infty$. Apart from that, some sensitive parameters are cancelled out in the process. It is easy to prove that under the same condition, most of the estimators proposed so far can achieve the same result, if not better,  as that proposed in \cite{DHPT06}.



In contrast to \cite{DHPT06}, \cite{ADV07, Zhu11a} propose an estimator that converts a general tree into a binary one and subsequently makes the likelihood equation into a quadratic equation of $A_k$  that is solvable analytically.   Experiments show the estimator preforms better than that in \cite{DHPT06} since the estimator uses more information in estimation.  Except experimental results, there is little statistical analysis to demonstrate why it is better than that proposed in \cite{DHPT06} and how to improve from there. Although the author of \cite{Zhu11a} proves the estimator is a MLE, there is a lack of other statistical properties, such as whether the MLE proposed in \cite{Zhu11a} is the same as that proposed in \cite{CDHT99} and if not, how much difference between them.

To be able to evaluate the performance of an estimator, we need to have the statistical properties of the estimator, such as unbiasedness, efficiency, variance, and so on, that differ from the asymptotic ones by showing the quality of an estimator in a finite sample. To distinguish the properties from the asymptotic ones, we call them finite sample properties and there has been a lack of results for the finite sample properties.  This paper aims to fill the gap and provides the properties.

\section{Assumption, Notation and Sufficient Statistics} \label{section2}

 To make the following statistical analysis clear and rigorous, we need to use  a large number of symbols that may overwhelm the readers who are not familiar with loss tomography. To assist them, the symbols  will be gradually introduced through the paper, where the frequently used symbols will be introduced in the following two sections and the others will be brought up  until needed. In addition, the most frequently used symbols and their meanings are presented in Table \ref{Frequently used symbols and description} for quick reference.

\subsection{Assumption}
We assume the probes multicasted from the source to receivers are
independent and network traffic remains statistically stable during the measurement. In addition, the observation obtained at receivers  is considered to be independent identical distributed
($i.i.d.$). Further, the losses occurred at a node or on a link are assumed to be $i.i.d$  as well.

\subsection{Notation}\label{treenotation}

As stated, a network of the tree topology is considered in this paper and denoted 
by $T=(V, E)$ that multicasts probes from the source to a number of receivers, where  $V=\{v_0, v_1, ... v_m\}$ is a set of
nodes and $E=\{e_1,..., e_m\}$ is a set of directed links that connect the
nodes in $V$. In addition, $v_k, k \in \{1,\cdot\cdot, m\}$ is often called node $k$ and $e_k$ called link $k$ in the following discussion. By default,  node $0$ is the root node of the multicast tree to which the source is attached. Apart from being the root that does not have a parent, node $0$ is different from  others by having a single descendant, $v_1$, that is connected by $e_1$.  Among the nodes in $V$, there are a number of them called leaf nodes that do not have any descendant but a receiver is attached to a leaf node. Because of this, we do not distinguish between a leaf node and a receiver and we use $R, R \subset V$ to denote them. Since there are $m$ links to connect  $m+1$ nodes in $T$, the links and nodes are organised in such a way that if $f(i)$ is used to denote
the parent of node $i$,  $e_i$ is the link connecting $v_{f(i)}$ to
$v_i$.  Figure \ref{tree example} is an example of a multicast binary tree that is named and connected according to the rules.

A multicast tree, as a tree, can be decomposed into a number of
 multicast subtrees at node $i, i \in V\setminus ( v_0 \lor R)$, where $T(i)$ denotes the multicast subtree that has $v_{f(i)}$ as its root, $e_i$ as its root link,  and $R(i)$ as the receivers attached to $T(i)$.  In addition, we use $d_i$  to denote the descendants attached to node $i$ that is a nonempty set if $i \notin R$. If $x$ is a set, $|x|$ is used to denote the number of elements in $x$. Thus, $|d_i|$ denotes the number of descendants in  $d_i$. Using the symbols on Figure \ref{tree example}, we have $R=\{v_8, v_9, \cdot\cdot, v_{15}\}$,  $R(v_2)=\{v_8, v_9, v_{10}, v_{11}\}$,  $d_{v2}=\{v_4, v_5\}$,  and $|d_{v_2}|=2$.

 If $n$ probes are sent from $v_0$ to $R$ in an experiment,
each of them gives rise of an independent realisation
of the passing (loss) process $X$. Let $X^{(i)}, i=1,...., n$ donate the $i-th$ process, where $x_k^i=1, k\in V$ if probe $i$
reaches $v_k$; otherwise $x_k^i=0$. The sample
$Y=(x_j^{(i)})^{i \in \{1,..,n\}}_{j \in R}$ is  the observation obtained in an experiment that can be divided into a number of sections according to $R(k)$, where $Y_k, k \in V$ denotes the part of $Y$ obtained by $R(k)$. In addition, each of the sections can be further divided into subsections $Y_x, x \subset d_k$ that is the part of observation obtained by $R(j), j \in x \land x \subset d_k$. Obviously, $Y_x \subset Y_k$.
If we use $y_j^i$ to denote the observation of receiver $j$ for probe $i$, we have $y_j^i=1$ if probe $i$ is observed by receiver $j$; otherwise, $y_j^i= 0$.

 Although loss tomography aims at estimate the loss rate of a link,  the pass rate of the path connecting $v_0$ to $v_k, k \in V$ is often used as the parameter  to be estimated. Let $A_k$ be the pass rate of the path connecting $v_0$ to $v_k$ that is defined as the percentage of the number of probes arrived at node $k$ among the number of probes sent by the source.  Given $A_k, k \in V\setminus v_0$, we are able to compute the pass rates of all links in $E$ since there is a bijection from the pass rates of the paths to the pass rates of the links in a network of the tree topology.  If $\alpha_k$ denotes the pass rate of link $k$ we have
 \begin{equation}
 \alpha_k=\dfrac{A_k}{A_{f(k)}}.
 \end{equation}
Given $\alpha_k$, we are able to compute the loss rate of link $k$ that is equal to $\bar{\alpha}_k=1-\alpha_k$.

\subsection{Statistics}

\label{mlesection}

To estimate  $A_k$ from $Y$, we need a likelihood function to connect the {\it i.i.d.} model defined previously to $Y$. To support the initiative of using a part of the available correlations to estimate $A_k$, a function, $n_k(x), x \subseteq d_k$, defined as follows is used to return the statistic for the likelihood function:
 \begin{equation}
 n_k(x)=\sum_{i=1}^n \bigvee_{\substack{j \in R(z)\\ z \in x}} y_j^i.
 \label{nk2}
 \end{equation}
Obviously.
\begin{equation}
 n_k(d_k)=\sum_{i=1}^n \bigvee_{\substack{j \in R(k)}} y_j^i.
 \label{nk1}
 \end{equation} \noindent  $n_k(x)$ is the number of probes, confirmed by the observation of $R(i), i \in x$, reaching node $k$. If $N_k$ is used to denote the number of probes reaching node $k$, we have $N_k \geq n_k(d_k)\geq n_k(x), x \subset d_k$, where $n_k(d_k)$ and $n_k(x)$ are of statistics that can be used to estimate $A_k$.

 To write a likelihood function of $A_k$ with $n_k(d_k)$, $\beta_k$ and $\gamma_k$ are introduced to denote the pass rate of the subtrees rooted at node $k$ and the pass rate of the special multicast tree that connects $v_0$ to node $k$ and then to $R(k)$. Clearly $\gamma_k=A_k\cdot\beta_k,  k \in V$ and $\hat\gamma_k=\dfrac{n_k(d_k)}{n}$ that is the empirical value of $\gamma_k$. Note that $\hat\gamma_j=\dfrac{n_j(j)}{n}, j \in R$ is the empirical pass rate of the path from the root to node $j$.
Given the assumptions and definitions,  the likelihood function of $A_k$ for observation $n_k(d_k)$ is written as follows:
\begin{equation}
{\cal L}(A_k, n_k(d_k))=(A_k\beta_k)^{n_k(d_k)}(1-A_k\beta_k)^{n-n_k(d_k)}.
\label{likelihood function}
\end{equation}
We can then prove  $n_k(d_k)$ is a sufficient statistic with respect to ({\it wrt.}) the passing process of $A_k$ for the observation obtained by $R(k)$. Rather than using the well known factorisation theorem in the proof, we directly use the mathematic definition of a sufficient statistic (See definition 7.18 in \cite{RM96}) to achieve this. The definition {\it wrt.} the statistical model defined for the passing process is presented as a theorem here:

\begin{figure}
\begin{center}
\begin{tikzpicture}[scale=0.25,every path/.style={>=latex},every node/.style={draw,circle,scale=0.8}]
  \node            (b) at (25,30)  { $v_0$ };
  \node            (d) at (25,24) { $v_1$ };
  \node            (f) at (20,18)  { $v_2$ };
  \node            (g) at (30.5,18) { $v_3$ };
  \node            (j) at (16,12)  { $v_4$ };
  \node            (k) at (23,12) { $v_5$ };
  \node            (l) at (29,12) { $v_6$ };
  \node            (m) at (35,12)  { $v_7$ };
  \node            (r) at (12,6)  { $v_8$ };
  \node            (s) at (18,6) { $v_9$ };
  \node            (t) at (21,6) { $v_{10}$ };
   \node            (u) at (24.5,6)  { $v_{11}$ };
  \node            (v) at (27.5,6)  { $v_{12}$ };
  \node            (w) at (30.5,6) { $v_{13}$ };
  \node            (x) at (33.5,6) { $v_{14}$ };
  \node            (y) at (39,6) { $v_{15}$ };

  \draw[->] (b) edge (d);
  \draw[->] (d) edge (f);
  \draw[->] (d) edge (g);
  \draw[->] (f) edge (j);
  \draw[->] (f) edge (k);
  \draw[->] (g) edge (l);
  \draw[->] (g) edge (m);
  \draw[->] (j) edge (r);
  \draw[->] (j) edge (s);
  \draw[->] (k) edge (t);
  \draw[->] (k) edge (u);
  \draw[->] (l) edge (v);
  \draw[->] (l) edge (w);
  \draw[->] (m) edge (x);
  \draw[->] (m) edge (y);
\end{tikzpicture}
\caption{A Multicast Tree} \label{tree example}
\end{center}
\end{figure}
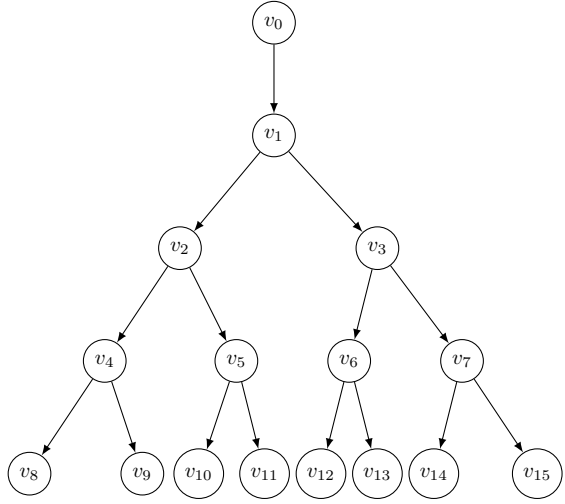

\begin{theorem}\label{complete minimal sufficient statistics}
Let $Y_k=\{X^{(1)},....,X^{(n)}\}$ be an i.i.d random sample, governed by  ${\cal L}(A_k|Y_k)$. The statistic $n_k(d_k)$ is minimal
sufficient for $A_k$ in respect of the observation of $Y_k$.
\end{theorem}

\begin{IEEEproof}
According to the definition of sufficiency, we need to prove
\begin{equation}
{\cal L}(A_k|n_k(d_k)=t)=\dfrac{{\cal L}(A_k, n_k(d_k)=t)}{{\cal L}(n_k(d_k)=t)}
\label{suff-condition}
\end{equation} is independent of $A_k$.

Given (\ref{likelihood function}), the passing process with observation of $n_k(d_k)=t$ is a random process that yields the binomial distribution as follows
\[
{\cal L}(n_k(d_k)=t)=\binom{n}{t}(A_k\beta_k)^{t}(1-A_k\beta_k)^{n-t}.
\]
Then, we have
\begin{eqnarray}
{\cal L}(A_k|n_k(d_k)=t)&=&\dfrac{(A_k\beta_k)^{t}(1-A_k\beta_k)^{n-t}}{\binom{n}{t}(A_k\beta_k)^{t}(1-A_k\beta_k)^{n-t}.
} \nonumber \\
=\dfrac{1}{\binom{n}{t}},
\end{eqnarray}
which is independent of $A_k$. Then, $n_k(d_k)$ is a sufficient statistic.

Apart from the sufficiency,
 $n_k(d_k)$,  as defined in (\ref{nk1}), is a count of the probes reaching $R(k)$ that counts each probe once and once only regardless of how many receivers observe the probe. Therefore,  $n_k(d_k)$ is a minimal sufficient statistic in regard to the observation of $R(k)$.
\end{IEEEproof}

\subsection{Statistics considering a part of observation}
\label{mlestatistics}
Instead of using $n_k(d_k)$ to estimate $A_k$, we can use $n_k(x), x \subset d_k \land |x|\geq 2$, defined in Section \ref{mlesection} to estimate $A_k$. The difference between them is the number of correlations considered in estimation, where the latter  is smaller than the former. As (\ref{likelihood function}), $\beta_k(x), x \subset d_k$ is needed to express the pass rate of the subtrees consisting of $T(j), j \in x$. Given $n_k(x)$ and $\beta_k(x)$, we can also write a likelihood function of $A_k$ and use the same procedure as that in Section \ref{mlesection} to prove $n_k(x)$ a sufficient statistic in the context of the observation obtained by $R(j), j \in x$. Further, an estimator on the observation of $R(j), j \in x$ can be created that will be discussed in Section \ref{section 4}.

\begin{table}[htdp]
\caption{Frequently used symbols and description}
\begin{center}
\begin{tabular}{|c|l|} \hline
Symbol & Desciption \\\hline
$T(k)$ & the subtree rooted at link $k$. \\ \hline
$d_k $& the descendants attached to node $k$. \\\hline
$R(k)$ & the receivers attached to $T(k)$. \\ \hline
$A_k$ & the pass rate of the path from $v_0$ to $v_k$. \\ \hline
$\beta_k$& the pass rate of the subtree rooted at node $k$. \\ \hline
$\beta_k(x)$& the pass rate of the subtree consisting of $T(j), j \in x \land x \subset d_k$. \\\hline
$\gamma_k$& $A_k*\beta_k$, pass rate from $v_0$ to $R(k)$. \\ \hline
$N_k$ & the number of probes reaching node $k$. \\ \hline
$x_k^i$ & the state of $v_k$ for probe $i$.  \\ \hline
$\sum_k$ & the $\sigma$-algebra created from $d_k$. \\ \hline
$n$ & the number of probes sent in an experiment, \\ \hline
$n_k(d_k)$ & the number of probes reaches $R(k)$. \\ \hline
$n_k(x)$ & the number of probes reaches the receivers attached to $T(j), j \in x$. \\ \hline
$I_k(x)$ & the number of probes observed by the members of $x$. \\ \hline
$Y$ &  the observation obtained in an experiment. \\ \hline
$Y_k, k \in V$& the part of $Y$ obtained by $R(k)$. \\ \hline
$Y_x, x \subset d_k$&  the part of $Y$ obtained by $R(j), j \in x$.\\ \hline
\end{tabular}
\end{center}
\label{Frequently used symbols and description}
\end{table}

\section{Estimator Analysis} \label{section3}
This section is dedicate to the analysis of the MLE that considers all of the correlations available in observation. By the analysis, we are able to identify all of the correlations in observation and find the connections among them that will set up the foundation for various explicit estimators.

\subsection{Maximum Likelihood Estimator based on $n_k(d_k)$} \label{2.a}

Turning the likelihood function presented in (\ref{likelihood function}) into a log-likelihood function, we have
 \begin{equation}
 \log {\cal L}(A_k|Y_k)=n_k(d_k)\log (A_k\beta_k)+(n-n_k(d_k))\log(1-A_k\beta_k).
 \label{likelihood}
 \end{equation}
 Differentiating (\ref{likelihood}) {\it wrt.} $A_k$
and letting the derivatives be 0, we have

\begin{eqnarray}
\dfrac{n_k(d_k)}{A_k}-\dfrac{(n-n_k(d_k))\beta_k}{1-A_k\beta_k}=0,
\label{likelihood equation}
\end{eqnarray}
and then
\begin{eqnarray}
A_k\beta_k&=&\frac{n_k(d_k)}{n}.
\label{AkBk}
\end{eqnarray}
Since neither $A_k$ nor $\beta_k$ can be solved from (\ref{AkBk}), we need to consider other correlations and then derive a MLE.
Given the {\it i.i.d.} model assumed previously and the multicast used in probing, we have the following equation to link the observation of $R(k)$ to $\beta_k$
\begin{equation}
1-\beta_k=\prod_{j \in d_k} (1-\dfrac{\gamma_j}{A_k}).
\label{beta-k}
\end{equation}
Solving $\beta_k$ from (\ref{beta-k}) and using it in (\ref{likelihood equation}), we have a MLE as
\begin{equation}
1-\dfrac{n_k(d_k)}{n \cdot A_k}=\prod_{j \in d_k} (1-\dfrac{\gamma_j}{A_k}).
\label{realmle1}
\end{equation}
Using $\gamma_k$ to replace $\dfrac{n_k(d_k)}{n}$ since the latter is the empirical value of the former, we have a likelihood equation as follows:
\begin{equation}
1-\dfrac{\gamma_k}{A_k}=\prod_{j \in d_k} (1-\dfrac{\gamma_j}{A_k})
\label{minc}
\end{equation}
that is identical to the estimator proposed in \cite{CDHT99}.

\subsection{Predictor and  Observation}
%
%

 To make the correlations involved in (\ref{realmle1}) visible,  we expand  the left hand side (LHS) and the right hand side (RHS) of  (\ref{realmle1}), where the terms obtained from the LHS are called observations and the terms from the RHS are called correlations. The correlation is also called  the {\it predictor} since it predicates the observation received in an experiment.  For instance, $\gamma_i\cdot \gamma_j/A_k, i, j \in d_k \land i \neq j$ is the predictor of the probes simultaneously observed by the receivers attached to subtree $i$ and subtree $j$, i.e. there is at least one receiver from each subtree. 

To represent the correlations involved in (\ref{realmle1}),
 a $\sigma$-algebra, $S_k$, is created over $d_k$ and  let $\Sigma_k=S_k \setminus \emptyset$ be the non-empty sets in $S_k$. Each member in $\Sigma_k$ corresponds to a pair of a predictor and its observation.  If  the number of elements in a member of $\Sigma_k$ is defined as the degree of the correlation, $\Sigma_k$ can be divided into $|d_k|$ exclusive groups, one for
 a degree of correlations that vary from 1 degree to $|d_k|$ degree. Let $S_k(i), i \in \{1,\cdot\cdot,|d_k|\}$ denote the group that considers $i$ degree correlations. For example, if $d_k=\{i,j,k,l\}$, $S_k(2)=\{(i,j),(i,k),(i,l),(j,k),(j,l),(k,l)\}$ consists of the pairwise correlations in $d_k$, and $S_k(3)=\{(i,j,k),(i,j,l),(i,k,l),(j,k,l)\}$ contains all of the triplet-wise correlations.

Given $\Sigma_k$,  $n_k(d_k)$ can be decomposed into the probes that are observed simultaneously by the members of  $\Sigma_k$ that is defined as if $x \in \Sigma_k$ and $|x|>1$, a probe observed by $x$  if and only if at least a receiver attached to subtree $j, j \in x$ observes the probe. We call such an observation simultaneous observation.
 To explicitly express $n_k(d_k)$ by $n_j(d_j), j \in d_k$,  $I_k(x), x \in \Sigma_k$ is introduced to return  the number of probes observed simultaneously by  $x$ in an experiment.
Let $u_j^i$ be the observation of $R(j)$ for probe $i$ that is defined as:
\[
u_j^i=\bigvee_{k \in R(j)} y_k^i,
\]
then
\begin{equation}
I_k(x)=\sum_{i=1}^n \bigwedge_{j \in x} u_j^i, \mbox{\vspace{1cm} } x \in \Sigma_k.
\label{I-k x}
\end{equation}
If $x=(j)$,
\[
I_k(x)=n_j(d_j), j \in d_k,
\]
Given the above, $n_k(d_k)$ can be decomposed as:

\begin{equation}
n_k(d_k)=\sum_{i=1}^{|d_k|}(-1)^{i-1}\sum_{x \in S_k(i)} I_k(x). \label{n_k value}
\end{equation}
(\ref{n_k value}) states that $n_k(d_k)$ is equal to a series of $I_k(x), x \in S_k(i)$ that are overlapped each other. To ensure each probe observed by $R(k)$ is counted once and once only in $n_k(d_k)$, we need to use the alternating adding and subtracting operations to eliminate duplication.

\subsection{Correspondence between Predictors and Observations}

Given  (\ref{n_k value}), we are able to prove the MLE proposed in \cite{CDHT99} considers all of the correlations in $\Sigma_k$ and have the following theorem.

\begin{theorem} \label{minctheorem}

\begin{enumerate}
\item (\ref{realmle1}) is a full likelihood estimator that considers all of the correlations in $\Sigma_k$;
\item (\ref{realmle1}) consists of  observed values and their predictors, one for a member of $\Sigma_k$; and
\item the estimate obtained from (\ref{realmle1})  is a fit that minimises an alternating differences between observed values and corresponding predictors.
    \end{enumerate}
\end{theorem}
\begin{IEEEproof}
(\ref{realmle1}) is a full likelihood estimator that considers all of the correlations in $d_k$. To prove 2) and 3),  we expand the both sides of (\ref{realmle1}) to pair the observed values with the  predictors of them according to $S_k$. We take three steps to achieve the goal.
\begin{enumerate}
\item  If  we use (\ref{n_k value}) to replace $n_k(d_k)$ from LHS of (\ref{realmle1}),  the LHS becomes:
\begin{equation}
1-\dfrac{n_k(d_k)}{n\cdot A_k}= 1-\dfrac{1}{n\cdot A_k}\big
[\sum_{i=1}^{|d_k|}(-1)^{i-1}\sum_{x \in S_k(i)}I_k(x)].
\label{nkexpansion}
\end{equation}
\item If we expand the product term located on  the RHS of (\ref{realmle1}), we have:
\begin{equation}
\prod_{j \in d_k}(1-\dfrac{\gamma_j}{A_k})=1-\sum_{i=1}^{|d_k|}(-1)^{i-1}\sum_{x \in S_k(i)}\dfrac{\prod_{j \in x} \gamma_j}{A_k^i} \label{prodexpansion}
\end{equation} where the alternative adding and subtracting operations intend to remove the impact of redundant observation.
\item Deducting 1 from both (\ref{nkexpansion}) and (\ref{prodexpansion})
and then multiplying the results by $A_k$, (\ref{realmle1}) turns to
\begin{eqnarray}
\sum_{i=1}^{|d_k|}(-1)^{i}\sum_{x \in S_k(i)}\dfrac{I_k(x)}{n}
=\sum_{i=1}^{|d_k|}(-1)^{i}\sum_{x \in S_k(i)}\dfrac{\prod_{j \in x} \gamma_j}{A_k^{i-1}}. \label{statequal}
\end{eqnarray}
It is clear there is a correspondence between the terms across
the equal sign, where the terms on the LHS are the observed values and the terms on the RHS are the predictors. If we rewrite
(\ref{statequal}) as
\begin{equation}
\sum_{i=1}^{|d_k|}(-1)^{i}\sum_{x \in S_k(i)}\Big(\dfrac{I_k(x)}{n} -\dfrac{\prod_{j \in x}
\gamma_j}{A_k^{i-1}}\Big)=0, \label{correspondence}
\end{equation}
\end{enumerate}
the correspondence between correlations and observed values becomes obvious.
\end{IEEEproof}

To distinguish the MLE from those proposed in this paper, we call it original MLE in the rest of the paper.

\section{Explicit Estimators based on Composite Likelihood} \label{section 4}
(\ref{correspondence}) shows that the original MLE takes into account all of the correlations in $\Sigma_k$. If the number of subtrees rooted at node $k$ is larger than 6, the estimator is a high degree polynomial that could not be solved analytically. To have an explicit estimator in those circumstances, we need to reduce the number of correlations considered in estimation and there are a number of strategies to achieve this. We here propose three of them and use composite likelihood that  is also called
pseudo-likelihood by Besag in \cite{Besay74} to create likelihood functions for the strategies. The three strategies are named  reduce scaled, block-wised, and individual based, respectively. The reduce scaled strategy, as named, is a small version of the original MLE that selectively removes a  number of subtrees rooted at node $k$ from consideration and then uses the maximum likelihood principle on the rest to estimate $A_k$. The block-wised strategy differs from the reduce scaled one by dividing all available correlations considered by the original MLE into a number of blocks, one for a degree of correlations, from pairwise to $d_k$-wise. The individual based one,  in contrast to the other two, considers a correlation at a time that leads to a large number of estimators.

\subsection{Reduce Scaled Estimator (RSE)}

Rather than considering all of the correlations in $\Sigma_k$, the correlations can be divided into groups according to the subtrees rooted at node $k$. Let $x, x \subset d_k$ be the group to be considered by an estimator in RSE. The log-likelihood function considering the correlations $x$ is as follows:
 \begin{equation}
\log L(A_k|Y_x)=n_k(x)\log (A_k\beta_k(x))+(n-n_k(x))\log(1-A_k\beta_k(x))
 \label{likelihood RSE}
 \end{equation}
 where $n_k(x)$ as defined in \ref{mlestatistics} is the number of probes reaching node $k$ confirmed from the observations of the receivers attached to $T(j), j \in x$ and
 $\beta_k(x)$ is the pass rate of  $T(j), j \in x$ that can be expressed as
\begin{equation}
1-\beta_k(x)=\prod_{j \in x} (1-\dfrac{\gamma_j}{A_k}).
\label{beta-k1}
\end{equation}
Then, a similar likelihood equation as (\ref{realmle1}) is obtained and presented as follows:
\begin{equation}
1-\dfrac{n_k(x)}{n \cdot A_k}=\prod_{j \in x} (1-\dfrac{\gamma_j}{A_k}).
\label{estimator MLEPC1}
\end{equation}
If $|x|<5$, the equation is solvable analytically. The estimators in RSE are denoted by $Am_k(x), x \subset d_k$.

\subsection{Block-wised Estimator (BWE)}

(\ref{correspondence}) shows that the correlations involved in the original MLE can be divided into $|d_k|-1$ blocks, from pairwise to $|d_k|$-wise.  Each of them can be written as a likelihood function. In order to use a unique likelihood function for all of them, we let the likelihood function considering single correlation be 1. Then, the $i$-wise likelihood function denoted as $L_c(i; A_k; y)$ can be expressed uniformly.

\begin{definition} \label{recursive corollary}
There are a number of composite likelihood functions, one for a degree of correlations, varying from pairwise to $|d_k|$-wise. The composite likelihood function  $L_c(i; A_k; y), i \in \{2,\cdot\cdot, |d_k|\}$ has a  form as follows:

\begin{eqnarray}
L_c(i; A_k; y) &=&\dfrac{\prod_{x \in S(i)} (A_k\beta_k(x))^{n_k(x)}(1-A_k\beta_k(x))^{n-n_k(x)}}{\prod_{x' \in S(i-1)}(A_k\beta_k(x'))^{n_k(x')}(1-A_k\beta_k(x'))^{n-n_k(x')}}. \nonumber \\
&& i \in \{2,\cdot\cdot,|d_k|\}
\label{recursive form}
\end{eqnarray}
 \end{definition}
 Let $A_k(i)$ be the estimator derived from $L_c(i; A_k; y)$. Then,  we have the following theorem.
\begin{theorem} \label{all explicit}
Each of the composite likelihood
equations obtained from (\ref{recursive form}) is an explicit estimator of $A_k$ that is as follows:
\begin{equation}
A_k(i)=\Big (\dfrac{\sum_{\substack{ x \in S_k(i)}}
\prod_{j \in x} \gamma_j}{\sum_{x \in S_k(i)}\dfrac{I_k(x)}{n}}{\Big )} ^{\frac{1}{i-1}},  i \in
\{2,.., |d_k|\}. \label{approximateestimator}
\end{equation}
\end{theorem}
\begin{IEEEproof}
Firstly, we can write (\ref{recursive form}) into a log-likelihood function and differentiate the log-likelihood function {\it wrt} $A_k$. As (\ref{AkBk}), we cannot solve $A_k$ or $\beta_k(x)$ directly from the derivative and we need to consider other correlations as (\ref{beta-k}). We then have an equation as
\begin{equation}
\frac{\partial\log L_c(i,A_k; y)}{\partial A_k}=\sum_{x \in
S(i)} \Big [1-\dfrac{\gamma_k(x)}{A_k}-\prod_{q\in x}(1-\dfrac{\gamma_q}{A_k})\Big ]-\sum_{x' \in S(i-1)} \Big [1-\dfrac{\gamma_k(x')}{A_k}-\prod_{q\in x'}(1-\dfrac{\gamma_q}{A_k})\Big ]
\label{pairwise equation}
\end{equation} The two summations can be expanded as (\ref{realmle1}) and only the terms related to the i-wise correlation left  since all other terms in the first summation are canceled by the terms of the second summation. The likelihood equation as (\ref{approximateestimator}) follows.
\end{IEEEproof}
In the rest of the paper, $A_k(i)$ is used to refer to the $i-wise$ estimator and $\widehat A_k(i)$ refers to the estimate obtained by $A_k(i)$.

\subsection{Individual based Estimator (IBE)}

Instead of considering a block of correlations together, we can consider a correlation at at time and have a large number of estimators. Each of them has a similar likelihood function as (\ref{likelihood RSE}), where $\beta_k(x)$ and $n_k(x)$ are replaced by $\psi_k(x)$ and $I_k(x)$, respectively. $\psi_k(x)=\prod_{j \in x} \alpha_j\beta_j, x \subseteq d_k$, is the pass rate of $T(j), j \in x$. If $\Sigma_k'=\Sigma_k \setminus S_k(1)$ is the correlations considered by IBE, the log-likelihood function for $A_k$ given observation $I_k(x)$ is equal to
 \begin{eqnarray}
 L(A_k|I_k(x))=I_k(x)\log (A_k\psi_k(x))+(n-I_k(x))\log(1-A_k\psi_k(x)),  \mbox{   } x \in \Sigma_k'.
 \label{Al likelihood1}
 \end{eqnarray}
 We then have the following theorem.
\begin{theorem}  \label{local estimator}
Given (\ref{Al likelihood1}), $A_k\psi_k(x)$ is a Bernoulli process. The MLE for $A_k$ given $I_k(x)$ equals to
\begin{equation}
Al_k(x)=\Big(\dfrac{\prod_{j\in x} \gamma_j}{\dfrac{I_k(x)}{n}}\Big)^{\frac{1}{|x|-1}}.  \mbox{   }    x \in \Sigma_k'
\label{local estimator1}
\end{equation}
\end{theorem}
	\begin{IEEEproof} Using the same procedure as that used in \ref{2.a}, we have the theorem.
\end{IEEEproof}

Comparing (\ref{approximateestimator}) with (\ref{local estimator1}), we can find that $\widehat Al_k(x)$, where $|x|=i$, is a type of geometric mean and $\widehat A_k(i)$ is the arithmetic mean of $\widehat Al_k(x), x \in S_k(i)$. Therefore, $A_k(i)$ is more robust than $Al_k(x)$.

Using and combining the strategies presented here, we can have various explicit estimators that cover those proposed previously. For instance, the estimator proposed in \cite{ADV07, Zhu11a} is one of them that divides $d_k$ into two groups and only considers the pairwise correlations between the members of the two groups. Therefore, although the estimator proposed in \cite{ADV07, Zhu11a} is a MLE in terms of the observation used in estimation, it is not the same as (\ref{minc}).

\section{Properties of the Estimators} \label{section5}

 It is known that if a MLE is a function of the sufficient statistic, it is asymptotically unbiased, consistent and asymptotically efficient. Thus, the original MLE and
all of the estimators proposed in this paper have the properties. Apart from them, we are interested in whether some of the estimators have better properties
 than them, such as, unbiasedness, uniqueness, variance, and efficiency, that can be used to evaluate the estimators. This section is devoted to present them that consist of a number of  theorems and corollaries.

\subsection{Unbiasedness and Uniqueness of $Al(x)$ and $A_k(i)$}
This subsection is focused on the unbiasedness of the estimators in IBE and BWE although the statistic used by the latter is not minimal sufficient. For $Al_k(x), x \in \Sigma_k'$, we have the following theorem.

 \begin{theorem} \label{local maximum}
$Al_k(x)$ is a unbiased estimator.
\end{theorem}

\begin{IEEEproof}
Let $z_j, j \in d_k$ be the pass rate of $T(j)$ and let  $\overline{A_k}=\frac{N_k}{n}$ be the sample mean of $A_k$. Note that $z_j$ and $z_l, j, l \in d_k$ are independent from each other if $ j \neq l$. In addition, $z_j, j \in d_k$ is independent from $A_k$. Because of this, $x_k^i\prod_{j \in x} z_j$ is used to replace $\bigwedge_{j \in x} y_j^i$ in the following derivation since the latter is equal to $\prod_{j\in x} y_j^i$ that is equal to $x_k^i\prod_{j \in x} z_j$. We then have

\begin{eqnarray}
E(\widehat Al_k(x))&=&E\Big( \big (\frac{\prod_{j\in x} \hat\gamma_j}{\frac{I_k(x)}{n}}\big)^{\frac{1}{|x|-1}}\Big) \nonumber \\
&=& E\Big(\big(\frac{\prod_{j\in x} \frac{n_j(d_j)}{n}}{\frac{\sum_{i=1}^n \bigwedge_{j \in x} y^i_j}{n}}\big)^{\frac{1}{|x|-1}} \Big )\nonumber \\
&=& E\Big(\big(\frac{(\dfrac{N_k}{n})^{|x|}\prod_{j\in x} \frac{n_j(d_j)}{N_k}}{\frac{N_k}{n} \frac{\sum_{i=1}^{N_k}\prod_{j \in x} z_j}{N_k}}\big)^{\frac{1}{|x|-1}} \Big ) \nonumber \\
&=& E\Big(\frac{N_k}{n}\Big) E\Big(\big(\frac{\prod_{j\in x} \frac{1}{N_k}\sum_{i=1}^{N_k}{z_j}}{\sum_{i=1}^{N_k} \frac{1}{N_k}\prod_{j \in x} {z_j}}\big)^{\frac{1}{|x|-1}}\Big ) \nonumber \\
&=& E\Big (\overline{A_k}\Big)
\end{eqnarray}

The theorem follows.
\end{IEEEproof}
Given theorem \ref{local maximum}, we have the follow corollary.
 \begin{corollary} \label{global expect}
 $A_k(i)$ is a unbiased estimator.
 \end{corollary}
    \begin{IEEEproof}
According to theorem \ref{local maximum}, we have
\begin{eqnarray}
E(\widehat A_k(i))&=&E\Big(\overline{A_k}\Big)E\Big(\big(\frac{\sum_{x \in S(i)}\prod_{j\in x} \frac{1}{N_k}\sum_{i=1}^{N_k}{z_j}}{\sum_{x \in S(i)}\sum_{i=1}^{N_k} \frac{1}{N_k}\prod_{j \in x} {z_j}}\big)\big)^{\frac{1}{i-1}} \Big)\nonumber \\
&=&E\Big (\overline{A_k}\Big)
\label{global estimate}
\end{eqnarray}
\end{IEEEproof}

%
%
%
Given $Al_k(x), x \in \Sigma_k'$ and $A_k(i)$ are unbiased estimators,
we can prove the uniqueness of $A_k(i)$.

\begin{theorem}
If
\[
\sum_{\substack{ x \in S_k(i)}} \prod_{j \in x} \hat\gamma_j < \sum_{x \in S_k(i)}\dfrac{I_k(x)}{n},
\]
there is only one solution in $(0,1)$ for $\widehat A_k(i), 2 \leq i \leq |d_k|$.
\end{theorem}
\begin{IEEEproof}
Since the support of $ A_k$ is in (0,1), we can reach this conclusion from (\ref{approximateestimator}).
\end{IEEEproof}

\subsection{Efficiency of  $Al_k(x)$, $Am_k(x)$, and the original MLE}
Apart from  asymptotically efficiency stated previously for the MLEs using sufficient statistics, we are interested in the efficiency of the estimators proposed in this paper. Given (\ref{Al likelihood1}), we have the following theorem for the Fisher information of an observation, $y$, for the estimators in IBE, i.e. $Al_k(x), x \in \Sigma_k'$.
\begin{theorem} \label{Al fisher}
The Fisher information of $y$ on $Al_k(x), x \subset d_k$ is equal to $ \dfrac{\psi_k(x)}{A_k (1-A_k \psi_k(x))}$.
\end{theorem}
\begin{IEEEproof}
Considering $I_k(x)=y$ is the observation of the receivers attached to $x$, we have the following as the likelihood function of the observation:
\begin{equation}
L(A_k|y)=y\log (A_k\psi_k(x))+(1-y)\log(1-A_k\psi_k(x)).
\label{Al likelihood for single}
\end{equation}
Differentiating (\ref{Al likelihood for single}) {\it wrt} $A_k$, we have
\begin{eqnarray}
\dfrac{\partial L(A_k|y)}{\partial A_k}=\dfrac{y}{A_k}-\dfrac{(1-y)\psi_k(x)}{1-A_k\psi_k(x)}
\end{eqnarray}
We then have
\begin{eqnarray}
\dfrac{\partial^2 L(A_k|y)}{\partial A_k^2}&=& -\dfrac{y}{A_k^2}-\dfrac{(1-y)\psi_k(x)^2}{(1-A_k\psi_k(x))^2}
\end{eqnarray}
If ${\cal I}(Al_k(x)|y)$ is used to denote the Fisher information of observation $y$ for $A_k$ in $Al_k(x)$, we  have
\begin{eqnarray}\label{fisher}
{\cal I}(Al_k(x)|y)&=&-E(\dfrac{\partial^2 L(A_k|y)}{\partial A_k^2}) \nonumber \\
&=&\dfrac{E(y)}{A_k^2}+\dfrac{E(1-y)\psi_k(x)^2}{(1-A_k\psi_k(x))^2} \nonumber \\
&=&\dfrac{\psi_k(x)}{A_k (1-A_k \psi_k(x))}
\end{eqnarray}
that is the information provided by $y$ for $A_k$.
\end{IEEEproof}
Given (\ref{fisher}), we are able to have a formula for the Fisher information of the original MLE and the estimators in RSE. In order to achieve this, let $\beta_k(d_k)=\beta_k$. Then, we have the following corollary.
\begin{corollary} \label{MLE fisher}
The Fisher information of observation $y$ for $A_k$ in the original MLE and $Am_k(x), x \subseteq d_k$ is equal to
\begin{equation}
 \dfrac{\beta_k(x)}{A_k (1-A_k \beta_k(x))}, \mbox{     } x \subseteq d_k.
 \label{MLE fisher equ}
 \end{equation}
\end{corollary}
\begin{IEEEproof}
Replacing $n_k(d_k)$ or $n_k(x)$ by $y$ and replacing $n-n_k(d_k)$ or $n-n_k(x)$ by $1-y$ from (\ref{likelihood}) and (\ref{likelihood RSE}), respectively, and then using the same procedure as that used in the proof of theorem \ref{Al fisher}, the corollary follows.
\end{IEEEproof}
Because of the similarity between (\ref{fisher}) and (\ref{MLE fisher equ}), the two equations have the same features in terms of support, singularity, and maximums. After
eliminating the singular points, the support of $A_k$ is in $(0,1)$ and the support of $\beta_k(x)$ (or $\psi_k(x)$) is in $[0, 1]$.
Both (\ref{fisher}) and (\ref{MLE fisher equ}) are convex functions in the support and reach the maximum at the points of  $A_k \rightarrow 1, \beta_k(x) =1$ (or ($\psi_k(x)=1$) and $A_k\rightarrow 0, \beta_k(x)=1$ (or ($\psi_k(x)=1$).
Given $A_k$, (\ref{MLE fisher equ}) is  a monotonic increase function of $\beta_k(x)$ whereas (\ref{fisher}) is a monotonic increase function of $\psi_k(x)$. 

Despite the similarity between (\ref{fisher}) and (\ref{MLE fisher equ}),  $Am_k(x)$ and $Al_k(x)$ react differently if $x$ is replaced by $y, x \subset y$ in terms of efficiency that leads to two corollaries, one for each of them.
\begin{corollary}
$Am_k(y)$ is more efficient than $Am_k(x)$ if $x \subset y$.
\end{corollary}
\begin{IEEEproof}
If $x \subset y$, $\beta_k(x) \leq \beta_k(y)$ and then we have the corollary.
\end{IEEEproof}
For $Al_k(x)$, we have
\begin{corollary} 
The efficiency of $Al_k(x), x \in \Sigma_k'$ forms a partial order that is identical to that formed on the inclusion of the members in $\Sigma_k'$, where the most efficient estimator must be one of the $Al_k(x), x \in S_k(2)$ and the least efficient one must be $Al_k(d_k)$.
\label{Al corollary}
\end{corollary}
\begin{IEEEproof}
According to Theorem \ref{Al fisher}, the efficiency of $Al_k(x)$ is determined by $\psi_k(x)$, where $\psi_k(x)=\prod_{j \in x} \alpha_j \beta_j$. If $x \subset y$, we have
 \begin{eqnarray}
 \psi_k(y)&=&\prod_{j \in y} \alpha_j \beta_j \nonumber \\
 &=&\psi_k(x)\prod_{j \in (y \setminus x)} \alpha_j \beta_j \nonumber \\
 &<& \psi_k(x)
 \end{eqnarray}.
 Therefore, the order of the efficiency of $Al_k(x), x \in \Sigma_k'$ shares that of the inclusion in $\Sigma_k'$, where $x, x \in S_k(2)$ are the members of $\Sigma_k'$ that have the minimal number of elements.
 In contrast to $\psi_k(x), x \in S_k(2)$, $\psi_k(d_k)\leq \psi_k(x)$ since $\forall x, x \in \Sigma_k', x \subseteq d_k$. Then, the corollary follows.
\end{IEEEproof}

\subsection{Variance of $Al_k(x)$, $Am_k(x)$, and the original MLE}
The estimator specified by (\ref{minc}),  $Am_k(x)$, and $Al_k(x)$ are of MLEs that have different focuses on the observations obtained. Despite the difference between them, they share a number of features, including likelihood function and efficient equation. In addition, the variances of them are expressed by a general function showing the connection between $A_k$ and the pass rate of the subtree considered in estimation. Let $mle$ denote all of them. Then, we have a theorem for the variances of the estimators in $mle$.

\begin{theorem} \label{Al variance}
The variances of the estimators in $mle$ equal to
\begin{equation}
var(mle)=\dfrac{A_k (1-A_k\delta_k(x) )}{\delta_k(x)}, \mbox{  } x \subseteq d_k
\label{Al variance1}
\end{equation}
where $\delta_k(x)$
\begin{eqnarray}
\delta_k(x)=\begin{cases} \beta_k(x), & \mbox{For original MLE and }  Am_k(x); \\  \psi_k(x), & \mbox{For } Al_k(x). \end{cases} \nonumber
\end{eqnarray}
\end{theorem}
\begin{IEEEproof}
The passing process described by (\ref{Al likelihood1}) is a Bernoulli process that falls into the exponential family and satisfies the regularity conditions presented in \cite{Joshi76}. Thus, the variance of  an estimator in $mle$ reaches the Cram\'{e}r-Rao bound that is the reciprocal of the Fisher information.
\end{IEEEproof}
(\ref{Al variance1}) can be written as
\begin{eqnarray}
\frac{A_k}{\delta_k(x)}-A_k^2
\end{eqnarray}
which shows:
\begin{enumerate}
 \item the estimates obtained by an estimator spread out more widely than that obtained by direct measurement. The wideness is determined by $\delta_k(x)$, the pass rate of the subtrees connecting node $k$ to observers. If $\delta_k(x)=1$, there is no further spread-out than that obtained by direct measurement. Otherwise, the variance increases as the decreases of $\delta_k$ and in a super linear fashion.
\item
the variance of the estimates obtained by an estimator is monotonically increasing as the depth of the subtree rooted at node $k$ since the pass rate of a subtree decreases as its depth, i.e., the pass rate of an $i$-level tree, say A, is larger than that of the $i+1$-level one that is extended from the $i$-level one;
\item the variance of the estimates  obtained by an estimator in $mle$ is a monotonically decreasing function of $\delta_k(x)$. 
\end{enumerate}
The three points confirm some of the experiment results reported previously, such as the dependency of variance on topology reported in \cite{CDHT99}. Note than despite 3), the variance of $Am_k(x)$ can be the same as that of $Am_k(y), x \subset y$ if $\beta_k(x)=\beta_k(y)$. So does $Al_k(x)$. In other words, if the probes observed by $R(j), j \in (y\setminus x)$ are included in that observed by $R(i), i \in x$, the estimate obtained by $Am_k(x)$ is the same as that obtained by $Am_k(y)$.

\subsection{Efficiency and Variance of BWE}

As stated, the estimate obtained by $A_k(i)$ is a type of the arithmetic mean of $Al_k(x), x \in S_k(i)$ that has the same advantages and disadvantages as the arithmetic mean. Thus, $A_k(i)$ is more robust and efficient than that of $Al_k(x), x \in S(i)$ since the former considers more probes than the latter in estimation although some of the probes may be considered more than once. Because of this, (\ref{fisher}) cannot be used to evaluate the efficiency of an estimator in BWE. Despite this, we can put a range for the information obtained by $A_k(i)$ that is
 \begin{eqnarray}
\frac{\psi_k(x)}{A_k(1-A_k\psi_k(x))} \leq {\cal I}(A_k(i)|y) \leq  {d_k \choose i} \frac{\psi_k(x)}{A_k(1-A_k\psi_k(x))}.  \mbox{    } |x|=i.
 \end{eqnarray}
In addition, $A(i)$ is at least as efficient as $A(i+1)$ and the variance of $A(i)$ is at least as small as that of $A(i+1)$ since $\sum_{x \in S_k(i)} I_k(x) \leq \sum_{x \in S_k(i+1)} I_k(x)$.

\subsection{Example}
We use an example to conclude this section that illustrate the differences of the variances obtained from the estimates of four estimators. The four estimators are: direct measurement, the original MLE, $Al_k(x), |x|=2$ and $Al_k(d_k)$, respectively. The setting used here is identical to that presented in \cite{DHPT06}, where node $k$ has three children with a pass
rate of $\alpha, 0 < \alpha \leq 1$, and the pass rate from the root to node $k$ is also equal to $\alpha$.  Using (\ref{Al variance1}), we have the variances of them that are presented below:
\begin{enumerate}
\item $\alpha-\alpha^2$,
\item $\frac{1}{3(1-\alpha)+\alpha^2}-\alpha^2$,
\item $\frac{1}{\alpha}-\alpha^2$, and
\item $\frac{1}{\alpha^2}-\alpha^2$.
\end{enumerate}
The difference between them becomes obvious as $\alpha$ decreases from $1$ to  $0.99$, where the variances of the four estimators change from 0 to 0.01, 0.01, 0.03, and 0.04, respectively. The variance of $Al_k(d_k)$ is 4 times of that of the original MLE that is significantly different from that obtained in \cite{DHPT06}. Although the variances are decreased as the number of probes multicasted, the ratio between them remains.

\section{Model Selection and Simulation} \label{section 6}
The large number of estimators in IBE, RSE and BWE, plus the original MLE, make model selection possible. However, to find the most suitable one in terms of efficiency and computational complexity is a hard task since the two goals conflict each other. Although one is able to identify the the most suitable estimator by computing the Kullback-Leigh divergence or the composite Kullback-Leigh divergence of the estimators, the cost of computing the Akaike information criterion (AIC) for each of the estimators makes this approach prohibitive.  Nevertheless, the derivation of (\ref{MLE fisher equ}) successfully solves the problem in some degree since (\ref{MLE fisher equ}) shows the most suitable estimator should have a bigger $\beta_k(x)$ which can be obtained from end-to-end observation since $\beta_k(x)  \propto \prod_{j \in x} \gamma_j$.

\begin{table*}[th]
  \centering
  \scriptsize
  \begin{tabular}{|l|r|r|r|r|r|r|r|r|r|r|}  \hline
 Estimators &\multicolumn{2}{|c|}{OMLE} & \multicolumn{2}{|c|}{$A_k(2)$} & \multicolumn{2}{|c|}{$A_k(3)$} &\multicolumn{2}{|c|}{$Al_k(x), |x|=2$} & \multicolumn{2}{|c|}{$Al_k(x), |x|=3$}	\\ \hline
samples & Mean & Var &	Mean & Var &	Mean & Var	& Mean & Var &  Mean &	Var	\\ \hline
300&	0.0088&	1.59E-05&	0.0088&	1.59E-05&	0.0088&	1.64E-05&	0.0087&	 1.59E-05&	0.0087&	1.61E-05 \\ \hline
900&	0.0092&	7.76E-06&	0.0092&	7.82E-06&	0.0091&	7.84E-06&	0.0092&	 7.90E-06&	0.0092&	8.15E-06 \\ \hline
1500&	0.0096&	4.55E-06&	0.0096&	4.55E-06&	0.0096&	4.80E-06&	0.0096&	 4.78E-06&	0.0096&	4.33E-06 \\ \hline
2100&	0.0097&	3.14E-06&	0.0097&	3.11E-06&	0.0097&	3.14E-06&	0.0097&	 3.02E-06&	0.0097&	3.08E-06 \\ \hline
2700&	0.0100&	1.72E-06&	0.0100&	1.72E-06&	0.0100&	1.74E-06&	0.0100&	 1.81E-06&	0.0100&	1.83E-06 \\ \hline
\end{tabular}
  \caption{Simulation Result of a 8-Descendant Tree with Loss Rate=$1\%$}
  \label{Tab2}
\end{table*}

\begin{table*}
\centering
\scriptsize
\begin{tabular}{|l|r|r|r|r|r|r|r|r|r|r|}  \hline
 Estimators &\multicolumn{2}{|c|}{OMLE} & \multicolumn{2}{|c|}{$A_k(2)$} & \multicolumn{2}{|c|}{$A_k(3)$} &\multicolumn{2}{|c|}{$Al_k(x), |x|=2$} & \multicolumn{2}{|c|}{$Al_k(x), |x|=3$}	\\ \hline
samples & Mean & Var &	Mean & Var &	Mean & Var	& Mean & Var &  Mean &	Var	\\ \hline
300&	0.0088&	1.59E-05&   0.0089&	1.64E-05&	0.0089&	1.68E-05&	0.0091&	 2.36E-05&	0.0088&	1.95E-05 \\ \hline
900&	0.0091&	7.76E-06&	0.0091&	7.80E-06&	0.0091&	7.83E-06&	0.0092&	 9.74E-06&	0.0091&	8.67E-06 \\ \hline
1500&	0.0096&	4.55E-06&	0.0096&	4.72E-06&	0.0096&	4.81E-06&	0.0097&	 4.36E-06&	0.0096&	4.45E-06 \\ \hline
2100&	0.0097&	3.14E-06&	0.0097&	3.11E-06&	0.0097&	3.11E-06&	0.0098&	 3.39E-06&	0.0097&	3.04E-06 \\ \hline
2700&	0.0100&	1.72E-06&	0.0100&	1.69E-06&	0.0100&	1.67E-06&	0.0101&	 2.11E-06&	0.0100&	1.90E-06 \\ \hline
\end{tabular}
  \caption{Simulation Result of a 8-Descendant Tree, 6 of the 8 have Loss Rate=$1\%$ and the other 2 have Loss Rate=$5\%$}
  \label{Tab3}
\end{table*}

\begin{table*}
\centering
\scriptsize
\begin{tabular}{|l|r|r|r|r|r|r|r|r|r|r|}  \hline
 Estimators &\multicolumn{2}{|c|}{OMLE} & \multicolumn{2}{|c|}{$A_k(2)$} & \multicolumn{2}{|c|}{$A_k(3)$} &\multicolumn{2}{|c|}{$Al_k(x), |x|=2$} & \multicolumn{2}{|c|}{$Al_k(x), |x|=3$}	\\ \hline
samples & Mean & Var &	Mean & Var &	Mean & Var	& Mean & Var &  Mean &	Var	\\ \hline
300&	0.0503&	2.15E-04&	0.0504&	2.15E-04&	0.0505&	2.14E-04&	0.0508&	2.18E-04&	0.0505&	2.16E-04\\ \hline
900&	0.0511&	5.85E-05&	0.0511&	5.81E-05&	0.0511&	5.79E-05&	0.0512&	5.79E-05&	0.0512&	5.88E-05\\ \hline
1500&	0.0502&	2.24E-05&	0.0502&	2.24E-05&	0.0502&	2.23E-05&	0.0503&	2.33E-05&	0.0502&	2.32E-05\\ \hline
2100&	0.0507&	1.16E-05&	0.0507&	1.19E-05&	0.0507&	1.20E-05&	0.0507&	1.09E-05&	0.0507&	1.13E-05\\ \hline
2700&	0.0507&	1.31E-05&	0.0507&	1.34E-05&	0.0507&	1.35E-05&	0.0508&	1.35E-05&	0.0507&	1.34E-05\\ \hline
\end{tabular}
  \caption{Simulation Result of a 8-Descendant Tree, the loss rate of the root link=$5\%$, 4 of the 8 have Loss Rate=$1\%$ and the other 4 have Loss Rate=$5\%$}
  \label{Tab4}
\end{table*}

\subsection{Simulation}

To compare the effectiveness, robustness, and sensitivity of the estimators between the original MLE, $A_k(i)$, and $Al_k(x)$, three rounds of simulations are conducted in three settings. The multicast tree used in the simulations having a path/link from the root to node $k$ that has 8 subtrees connecting to the receivers. Five estimators: the original MLE (OMLE), $ A_k(2), A_k(3), Al_k(x), |x|=2$,  and $Al_k(x), |x|=3$, are compared against each other in the simulation. The number of samples used in the simulations varies from 300 to 2700 in a step of 600. For each sample size, 20 experiments with different initial seeds are carried out and the means and variances of the estimates obtained by the five estimators are presented in three tables, from Table \ref{Tab2} to Table \ref{Tab4}.

 Table \ref{Tab2} is the results obtained in the first round that sets the loss rate of the subtrees to 1\%, so does the loss rate of the path from the root to node $k$.  The result shows when the sample size is small, the estimates obtained by all estimators are drifted away from the true value that indicates the data obtained is not enough. With the increase of sample size, the estimates gradually approach to the true value and all of the estimators achieve a  similar outcome. As expected, the variances decrease as the sample size that is agreed with (\ref{Al variance1}) and there is no significant difference among the estimators since all of the subtrees connected to node $k$ have the same loss rates. Despite this, the variance of $Al_k(x), |x|=2$ is slightly better than that of $Al_k(x), |x|=3$ as specified by Theorem \ref{Al variance}.

To compare the sensitivity and robustness, another round simulation is carried out on the same network.
 The difference between this round and the previous one is the loss rates of the subtrees connected to node $k$, where 6 of the 8 subtrees have their loss rates equal to $1\%$ and the other two have their loss rates equal to $5\%$. The two subtrees
  considered by $Al_k(x), |x|=2$ have the loss rates equal to $1\%$ and $5\%$, respectively; whereas the two of the three subtrees considered by $Al_k(x), |x|=3$ have their loss rates equal to $1\%$ and the other has its loss rate equals to $5\%$. The results are presented in Table \ref{Tab3}. Compared Table \ref{Tab3} with Table \ref{Tab2}, there is no change for the original MLE and there are slight changes for the estimates obtained by $A_k(2)$ and $A_k(3)$. That confirms the robustness of the original MLE and the $A_k(i)$ over $Al_k(x)$. In contrast to the original MLE and $A_k(i)$, the variances of the estimates obtained by $Al_k(x), |x|=2$ and  $Al_k(x), |x|=3$ have noticeable differences with their counterparts, in particular  if the sample size is small because each of the estimators has a descendant with a higher loss rate than that used in the first round.  The advantage of this shows at the mean obtained by $Al_k(x), |x|=2$ that
 approaches to the true value quicker than that in the first round and that of $Al_k(x), |x|=3$.  This is because one of the two descendants considered by $Al_k(x), |x|=2$ has a higher loss rate than the other that increases the probability of  matching the predicator to its observation. In contrast to $Al_k(x), |x|=2$, the mean of $Al_k(x), |x|=3$ has little change from that obtained in the first round. This reflects the tradeoff between efficiency and robustness among $Al_k(x)$, where the larger the $|x|$ is, the robuster the $Al_k(x)$ is to the turbulence of the loss rates in $x$. 
 To have a similar result as the original MLE, we should select  the subtrees that have loss rates equating to $1\%$ for $Al_k(x), |x| = 2$ or $3$. Then, the same result as that presented in Table \ref{Tab2} should be obtained.

To verify the claim made at the end of last paragraph, we  conduct another round simulation, where the loss rate of the path of interest is increased from $1\%$ to $5\%$, and the loss rates of the eight subtrees rooted at node $k$ are divided into two groups, four of them are set to $5\%$ and the other four to $1\%$.  The two estimators from IBE, i.e. $Al_k(x), |x|=2$ and $3$, consider the observations obtained from the subtrees that have their loss rates equal to $1\%$. The result is presented in Table \ref{Tab4} that confirms the estimates of $Al_k(x)$ can be as good as that of the OMLE.  In comparison with
Table \ref{Tab2}, there are two noticeable differences in Table \ref{Tab4} :
 \begin{itemize}
 \item the means of the estimates approach to the true value quicker; and
 \item the  variances are a magnitude higher.
  \end{itemize}
The first can be derived from Theorem \ref{Al fisher}  and Corollary \ref{MLE fisher} since the efficiency of an estimator is inversely proportional to $A_k$; whereas the second can be obtained from  Theorem \ref{Al variance} that states a smaller $\delta_k(x)$ results in a bigger variance.

The simulations show that the original MLE undoultly is the most robust estimator that fits all of the three situations well although it reacts slower than some of the estimators proposed in this paper to the variation of observation. In contrast,  there is always an estimator that has a similar performance as that of the MLE in each of the situations. 
 The findings of this paper make it  possible to identify a suitable estimator according to end-to-end observation.

\section{Conclusion}\label{section7}

This paper starts from finding inspirations that can lead to efficient explicit estimators for loss tomography and ends with a large number of unbiased or asymptotic unbiased and consistent explicit estimators, plus a number of theorems and corollaries to assure the statistical properties of the estimators. One of the most important findings is of the formulae to compute the variances of $A_k$ estimated by the estimators in RSE, IBE and the original MLE. Apart from clearly expressing the connection between the path to be estimated and the subtrees connecting the path to the observers of interest, the formulae potentially have many applications in network tomography, some have been identified in this paper.  For instance, using the formulae, we have ranked the MLEs proposed so far, including those proposed  in this paper. In addition, the formulae make model selection possible in loss tomography and then the multicast used in end-to-end measurement is no longer only for creating various correlations but also for identifying the subtrees that can be used in estimation. The effectiveness of the strategy has been verified in a simulation study. Apart from those, there are other potentials to use the formulae and the findings that require further exploration.


\bibliography{../globcom06/congestion}

\begin{thebibliography}{10}

\bibitem{YV96}
Y.~Vardi, ``Network tomography: Estimating source-destination traffic
  intensities from link data,'' {\em Journal of Amer. Stat. Association},
  vol.~91, no.~433, 1996.

\bibitem{CDHT99}
R.~C\'{a}ceres, N.~Duffield, J.~Horowitz, and D.~Towsley, ``Multicast-based
  inference of network-internal loss characteristics,'' {\em IEEE Trans. on
  Information Theory}, vol.~45, 1999.

\bibitem{CDMT99}
R.~C\'{a}ceres, N.~Duffield, S.~Moon, and D.~Towsley, ``{Inference of Internal
  Loss Rates in the MBone },'' in {\em IEEE/ISOC Global Internet'99}, 1999.

\bibitem{CDMT99a}
R.~C\'{a}ceres, N.~Duffield, S.~Moon, and D.~Towsley, ``Inferring link-level
  performance from end-to-end multicast measurements,'' tech. rep., University
  of Massachusetts, 1999.

\bibitem{CN00}
M.~Coates and R.~Nowak, ``Unicast network tomography using {EM} algorthms,''
  Tech. Rep. TR-0004, Rice University, September 2000.

\bibitem{XGN06}
B.~Xi, G.~Nichailidis, and V.~Nair, ``Estimating netwrok loss rates using
  active tomography,'' {\em Journal of the American Statistical Association},
  vol.~101, no.~476, 2006.

\bibitem{BDPT02}
T.~Bu, N.~Duffield, F.~Presti, and D.~Towsley, ``Network tomography on {General
  Topologies},'' in {\em SIGCOMM 2002}, 2002.

\bibitem{ADV07}
V.~Arya, N.~Duffield, and D.~Veitch, ``Multicast inference of temporal loss
  charateristics,'' {\em Performance Evaluation}, vol.~9-12, 2007.

\bibitem{DHPT06}
N.~Duffield, J.~Horowitz, F.~L. Presti, and D.~Towsley, ``Explicit loss
  inference in multicast tomography,'' {\em IEEE trans. on Information Theory},
  vol.~52, no.~8, 2006.

\bibitem{ZG05}
W.~Zhu and Z.~Geng, ``Bottom up inference of loss rate,'' {\em Journal of
  Computer Communications}, vol.~28, no.~4, 2005.

\bibitem{GW03}
D.~Guo and X.~Wang, ``Bayesian inference of network loss and delay
  characteristics with applications to tcp performance predication,'' {\em IEEE
  trans. on Signal Processing}, vol.~51, no.~8, 2003.

\bibitem{LY03}
G.~Liang and B.~Yu, ``Maximum pseudo likelihood estimation in network
  tomography,'' {\em IEEE trans. on Signal Processing}, vol.~51, no.~8, 2003.

\bibitem{TCN03}
Y.~Tsang, M.~Coates, and R.~D. Nowak, ``Network delay tomography,'' {\em IEEE
  trans. on Signal Processing}, vol.~51, no.~8, 2003.

\bibitem{PDHT02}
F.~L. Presti, N.~Duffield, J.~Horowitz, and D.~Towsley, ``Multicast-based
  inference of network-internal delay distributions,'' {\em IEEE/ACM trans. on
  Networking}, vol.~10, 2002.

\bibitem{SH03}
M.-F. Shih and A.~Hero, ``Unicast-based inference of network link delay
  distribution with finite mixture models,'' {\em IEEE trans. on Signal
  Processing}, vol.~51, no.~8, 2003.

\bibitem{LGN06}
E.~Lawrence, G.~Michailidis, and V.~N. Nair, ``Network delay tomography using
  flexicast experiments,'' {\em Journal of Royal Statistica Society, Series B},
  vol.~68, no.~5, 2006.

\bibitem{HBB00}
K.~Harfoush, A.~Bestavros, and J.~Byers, ``Robust identification of shared
  losses using end-to-end unicast probes,'' in {\em Technical Report
  BUCS-2000-013}, Boston University, 2000.

\bibitem{Lindsay88}
B.~C. Lindsay, ``Composite likelihood method,'' {\em Contemporary Mathematics},
  vol.~80, 1988.

\bibitem{Zhu11a}
W.~Zhu, ``An efficient loss rate estimator in multicast tomography and its
  validity,'' in {\em IEEE International Conference on Communicaation and
  Software}, 2011.

\bibitem{RM96}
R.~Mittelhammer, {\em Mathematical Statistics for Economics and Business},
  vol.~78.
\newblock Springer, 1996.

\bibitem{Besay74}
J.~Besag, ``Spatial interaction and the statistical analysis of lattice system
  (with discussion),'' {\em Journal of Royal Statistical Society}, vol.~36,
  1974.

\bibitem{Joshi76}
V.~M. Joshi, ``On the attainment of the cramer-rao lower bound,'' {\em Ann.
  Statist.}, vol.~4, no.~5, pp.~998--1002, 1976.

\end{thebibliography}


\end{document}